\documentclass{article}
\usepackage[modulo,switch]{lineno}
\usepackage{multirow}
\usepackage[utf8]{inputenc}
\usepackage{amsmath}
\usepackage{amssymb}
\usepackage{graphicx}
\usepackage{authblk}

\makeatletter\@ifundefined{date}{}{\date{}}
\makeatother

\evensidemargin\oddsidemargin \hoffset-22.4mm \voffset-28.4mm
\title{Connected but Segregated: Social Networks in Rural Villages}
\author[1]{Felipe Montes}
\author[1]{Roberto C. Jimenez}
\author[2]{Jukka-Pekka Onnela}
\affil[1]{Department of Industrial Engineering, Universidad de los Andes, Social and Health Complexity Center, Bogotá, Colombia. fel-mont@uniandes.edu.co}
\affil[2]{Department of Biostatistics, Harvard T.H. Chan School of Public Health, Boston, MA 02115. onnela@hsph.harvard.edu}
\begin{document}

\maketitle\thispagestyle{empty}
\begin{abstract}
There is an increased appreciation for, and utilization of, social networks to disseminate various kinds of interventions in a target population. Homophily, the tendency of people to be similar to those they interact with, can create within-group cohesion but at the same time can also lead to societal segregation. In public health, social segregation can form barriers to the spread of health interventions from one group to another. We analyzed the structure of social networks in 75 villages in Karnataka, India, both at the level of individuals and network communities. We found all villages to be strongly segregated at the community level, especially along the lines of caste and sex, whereas other socioeconomic variables, such as age and education, were only weakly associated with these groups in the network. While the studied networks are densely connected, our results indicate that the villages are highly segregated.
\end{abstract}

\section{Introduction}

The study of social network structure has enabled the identification of social relations as conduits for the spread of health related behaviors in both randomized and observational studies~\cite{Centola2010,Christakis2007,Valente2010}. Several studies have now demonstrated the utility of social networks for identifying initial spreaders within networks and how they may be harnessed to increase the efficiency of public health and development interventions~\cite{Banerjee2013,Christakis2010,Hunter2015,Kim2015}. It has been shown that the community structure of networks, where a community refers to a set of densely interconnected nodes, is important for targeting public health interventions, especially interventions that may have spillover effects, i.e., when the effect of an intervention may spread from one person to another~\cite{Centola2011,Borge-Holthoefer2013}. An additional consideration is the role of homophily, the tendency for individuals to be connected to others like them, which can result in overly optimistic estimates of the effectiveness of different seeding strategies of interventions if not properly taken into account~\cite{Aral2013}.
Karnataka is a southern state of India with approximately 55 million inhabitants, and it agglomerates individuals from heterogeneous castes under a common language and religion. Prior research has studied the Karnataka networks at the household level with the goal of identifying injection points for a microfinance program~\cite{Banerjee2013}. The network structure within these villages would be expected to be influenced by the ancient caste system of social stratification that generates hierarchies, restricts dietary and social interactions, and creates physical and educational separation between people from different castes~\cite{Berreman1972}. Our goal in this paper is to carry out a detailed investigation into the structural properties of these networks, their community structure, and the role that nodal covariates play in giving rise to social segregation. A deeper understanding of these networks could make it possible to develop more effective intervention strategies that overcome the existing social and cultural barriers in villages similar to these throughout the world in resource poor settings.

\section{Results}
\subsection{Network Characteristics}\label{subs_net}

We used data, now available in the public domain, collected in 75 villages in Karnataka, India, in a study conducted by the Abdul Latif Jameel Poverty Action Lab in 2006 ~\cite{Banerjee2013}. As in the Diffusion of Microfinance study, for every village we constructed undirected single-layered networks, where a node represents an individual and an undirected tie connects two nodes if one of the individuals reported at least one of the 12 types of relationships with the other ~\cite{Banerjee2013} (see Methods). We also make use of the demographic and social network data collected from surveyed individuals. Nodal attributes or covariates include sex, age, religion (Hinduism, Islam, Christianity), caste (Scheduled Caste, Scheduled Tribe, Other Backward Class and General), level of education (years), an indicator for whether the individual worked during the week preceding the survey, and an indicator for whether the individual had a bank account. The networks sizes across villages vary from 354 to 1773 nodes with a median value of 869 (see Supplementary Materials).


\begin{table}[ht]
\centering
\caption{Characteristics of the 75 networks and of their largest connected components (LCCs). Here $N$ is the number of network nodes, $n$ the number of LCC nodes, $M$ the number of edges, $m$ is the number of LCC edges, $\delta$ the edge density, $\langle k \rangle$ the mean degree, and $\langle C \rangle$ the mean (local) clustering coefficient.}
\label{T_one}
\resizebox{\textwidth}{!}{
\begin{tabular}{c|c|c|c||c|c|c|}
\cline{2-7}
       & \multicolumn{3}{c||}{Networks}   &  \multicolumn{3}{c|}{Largest Connected  Components }  \\ \cline{2-7}

   &  Min   & Median   & Max   & Min   & Median   & Max    \\
\hline 
\multicolumn{1}{|c|}{$N$}   &  354   & 869   & 1773   & 346   & 850   & 1729     \\
\hline
\multicolumn{1}{|c|}{$M$}   &  1540   & 3750   & 7854   & 1519   &  3703  & 7818   \\
\hline
\multicolumn{1}{|c|}{$\delta$}    &  $4.9\times10^{-3}$   & $9.7\times10^{-3}$   & $2.4\times10^{-2}$   & $5.0\times10^{-3}$   & $1.0\times10^{-2}$   & $2.6\times10^{-2}$     \\
\hline
\multicolumn{1}{|c|}{$\langle k \rangle$}   &  6.8   & 8.4   & 10.4   & 7.0   & 8.6   & 10.6     \\
\hline
\multicolumn{1}{|c|}{$\langle C \rangle$}  &  0.6   & 0.6   & 0.7   & 0.6   & 0.6  & 0.7     \\
\hline
\multicolumn{1}{|c|}{Components}  &  1  & 7 & 25   & 1   &  1   & 1  \\
\hline
\multicolumn{1}{|c|}{$n / N$}   &  -   &  -   & -   & 89\%   & 98\%   & 100\% \\
\hline
\multicolumn{1}{|c|}{$m / M$}   &  -   & -   & -  & 95\%   & 99\%   & 100\%    \\
\hline
\end{tabular}}
\end{table}

For our analyses we use the largest connected components (LCCs) of the networks. All networks, with one exception, have more than one connected component and the LCCs of these networks contain a median value of 98\% of network nodes and 99\% of network ties (Table~\ref{T_one}). Moreover, the average degree and mean clustering coefficient of the LCCs are within 3\% of those computed for the full networks.

\subsection{Dyad-level assortativity}
\begin{figure}[tp]
\centering\includegraphics[width=6in]{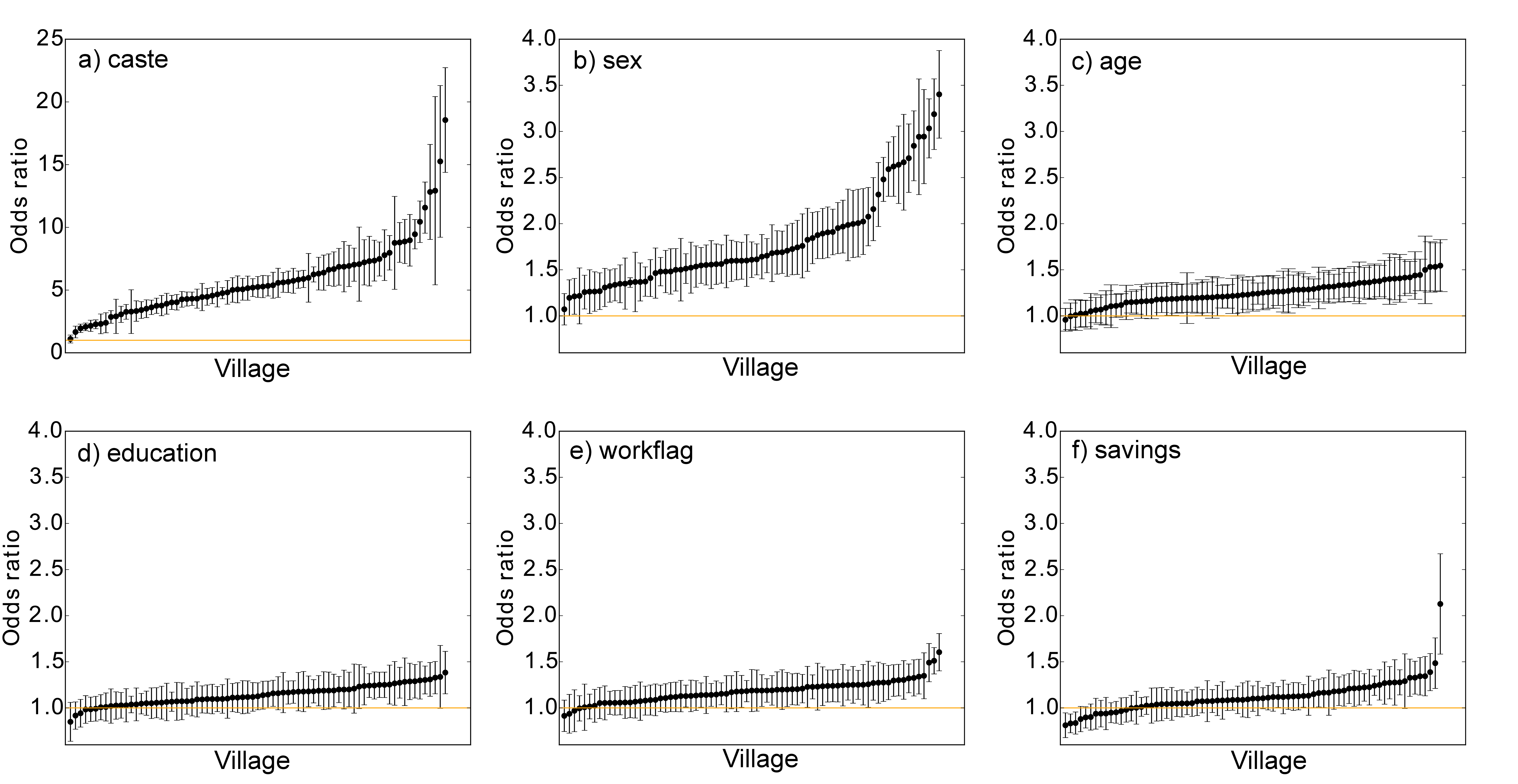}
\caption{Dyad-level assortativity odds ratios for different nodal attributes in each of the 75 LCCs. The error bars indicate 95\% confidence intervals for the odds ratios. Villages are ordered along $x$-axis in increasing order of odds ratios (point estimates), not according to village indices. The ordering of villages is consequently different in each panel.} 
\label{fig-one}
\end{figure}

Social segregation might be attributed to dyad-level assortativity which quantifies the extent to which pairs of connected nodes share the value of an attribute of interest ~\cite{Newman2003,Noldus2014}. By using a logistic regression model, which assumes dyadic independence, we modeled the existence of a tie between a pair of nodes based on the sex, age, caste, religion, education, employment and savings of the individuals. We found that, at the dyadic level, 98\% of the networks had significant assortativity based on the caste attribute (Figure~\ref{fig-one}). Across the villages, individuals belonging to the same caste are 1.10 to 18.56 times more likely to form a tie than individuals belonging to different castes. Assortativity based on age, savings, work and education attributes was significant in 80\%, 44\%, 41\%, and 60\% of the villages, respectively, and the odds ratios for these attributes ranged from 0.96 to 1.54 for age, 0.81 to 2.13 for savings, 0.91 to 1.60 for work flag and 0.85 to 1.38 for education (Table~\ref{T_two}). These results show that connected node pairs are much more similar in terms of their nodal attributes than unconnected node pairs. Nevertheless, these associations remain weaker than that associated with the caste attributes of individuals. In addition, assortativity for these attributes are similar and vary in a small range across the villages. In the 75 LCCs, 95.5\% of individuals belong to the same religion, and consequently assortativity associated with religion was significant only in 38\% of the networks with odds ratios lower than 1.5 except for one village where the odds ratio is close to 2. By contrast, individuals with the same sex are 1.1 to 3.4 times more likely to form a tie than individuals of the opposite sex (odds-ratio cumulative distributions are available in the supplementary materials).

\begin{table}[ht]
\centering
\caption{Dyad-level assortativity odds ratios and community-based assortativity mutual information coefficient $\tilde{I}_d$ for different nodal attributes in each of the 75 LCCs}
\label{T_two}
\resizebox{\textwidth}{!}{
\begin{tabular}{|c|c|c|c|c|c|c|c|c|}
\cline {1-7}
  Attribute & Number of & LCCs with & \multicolumn{3}{c|}{Dyad-level} & Mutual \\ 
  & categories & significant dyad-level & \multicolumn{3}{c|}{odds-ratio} & Information \\
  &  & assortativity (\%)  & \multicolumn{3}{c|}{odds-ratio} & coefficient\\
  \cline{1-7}
  & & & Min & Median & Max & \\
\hline 
\multicolumn{1}{|c|}{Caste} & 4 & 99 & 4.31 & 5.06 & 18.56 & 0.39 \\
\hline
\multicolumn{1}{|c|}{Sex} & 2 & 97 & 1.10 & 1.56 & 3.40 & 0.01 \\
\hline
\multicolumn{1}{|c|}{Age} & 6 & 80 & 0.96 & 1.22 & 1.54 & 0.08 \\
\hline
\multicolumn{1}{|c|}{Workflag} & 2 & 41 & 0.91 & 1.15 & 1.60 & 0.40 \\
\hline
\multicolumn{1}{|c|}{Education} & 6 & 60 & 0.85 & 1.11 & 1.38 & 0.10 \\
\hline
\multicolumn{1}{|c|}{Savings} & 2 & 44 & 0.81 & 1.09 & 2.13 & 0.05 \\
\hline
\multicolumn{1}{|c|}{Religion} & 4 & 38 & 0.49 & 1.05 & 48.52 & 0.08 \\
\hline
\end{tabular}}
\end{table}

The logistic model results for dyad-level assortativity by sex do not reflect whether the effect is attributed to male-male or female-female relations. To further investigate dyadic assortativity by sex, we complemented the logistic model results by formulating a mean degree constrained null model and determining whether male-male, female-female and female-male relations were associated with preferential tie formation among individuals. This statistical test is based on keeping the structure of the network fixed and randomly reassigning the sex attribute for each node by a random permutation. According to the results, male-male ties occur more frequently than expected by chance in 72 (96.0\%) villages (Table~\ref{T_three}). Sex dissortativity is rare and, in fact, the male-female relations were less likely to occur than expected by chance in only 2 networks (2.7\%). Female-female ties were more common than expected by chance in just 8 (10.7\%) of the villages, in the other 67 (89.3\%) villages the test results were not statistically significant. As a sensitivity analysis, we also report the results when the mean degrees of males and females are allowed to deviate as much as 20\% from their empirically observed counterparts. As a consequence, the percentage of networks with male assortativity decreases to 74\%, the percentage of networks with female assortativity increases to 14\%, and the percentage of networks where female-male ties are less likely to occur than by chance remains at 97.4\%. This shows that with relaxed restrictions on the permutation there is still male-male preference  in most of the cases and female-female preference in some cases. Sex dissortativity results are unaffected by relaxation of the null model.

\begin{table}[ht]
\centering \caption{Results for dyad-level assortativity and dissortativity based on sex for 75 villages (networks) in Karnataka.}\label{T_three}
\vspace{1mm}
\begin{tabular}{|c|c|c|c|}
\hline
  Degree & Relationship & Percentage of & Percentage of\\
  tolerance & type &villages with & villages with\\
   &   & assortativity & dissortativity\\
\hline
\multirow{3}{*}{5\%}   &  Male-Male   & 96.0\%   & 0.0\% \\
\cline{2-4}
       &  Male-Female   & 0.0\%   & 97.3\% \\
\cline{2-4}
   &  Female-Female   & 10.7\%   & 0.0\% \\
\hline
\multirow{3}{*}{20\%}    &  Male-Male  &  76\%  &  0.0\% \\
\cline{2-4}
       &  Male-Female   & 0.0\%   & 97.3\% \\
\cline{2-4}
        &  Female-Female    & 16.0\%   & 0.0\% \\
\hline

\end{tabular}
\end{table}

\subsection{Community-level Assortativity}
While dyadic analysis permits us to quantify assortativity at the level of node pairs, it does not provide evidence of assortativity at the level of groups consisting of more than two nodes. We assess community-level assortativity by first detecting network communities ~\cite{Kumpula2007,Fortunato2010,Porter2009} and then investigating the extent to which network communities share nodal attribute values. We apply modularity maximization ~\cite{Newman2003,Girvan2002} to detect communities in the LCC of each network using the so-called Louvain heuristic for maximizing modularity~\cite{Blondel2008} (For information on the number of communities per LCC and the communities size distribution see Supplementary Materials). In Figure~\ref{fig-two} we provide intuition about which attributes might be associated with network communities detected. We visualized the village networks with nodes colored by the network community assignment, sex, age, religion, caste, education and employment indicators and savings. We show these visualizations for one of the villages (village 52) and observe that caste appears most strongly associated with network communities.  

\begin{figure}[tp]
\centering\includegraphics[width=5in]{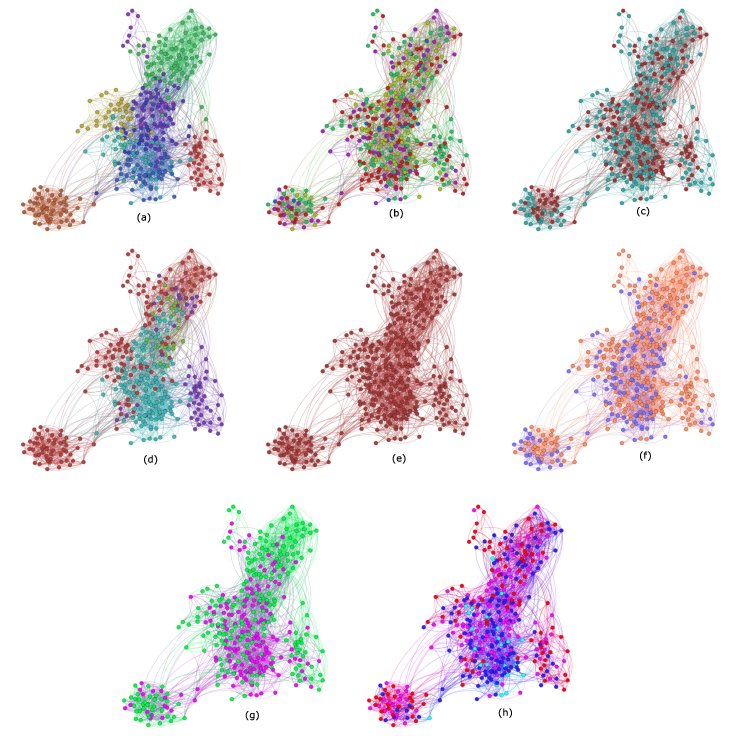}
\caption{Visualization of a village social network colored by (a) community assignment (obtained via modularity maximization), (b) age (purple: 18-30 years, red: 31-40 years, green: 41-50 years, light green: 51-64 years, blue: $>64$ years), (c) sex (red: male, blue: female), (d) caste (purple: scheduled caste, green: scheduled tribe, red: OBC, cyan: general), (e) religion (red: Hinduism), (f) work flag (orange: worked last week, purple: did not work last week), (g) savings (purple: does not have a bank account, green: has a bank account), (h) education (pink: 1-9 years, blue: 10-13 years, cyan: 14-15 years, red: no education). For all panels, nodes with missing covariates are not visualized.}
\label{fig-two}
\end{figure}

\begin{figure}[tp]
\centering\includegraphics[width=5.5in]{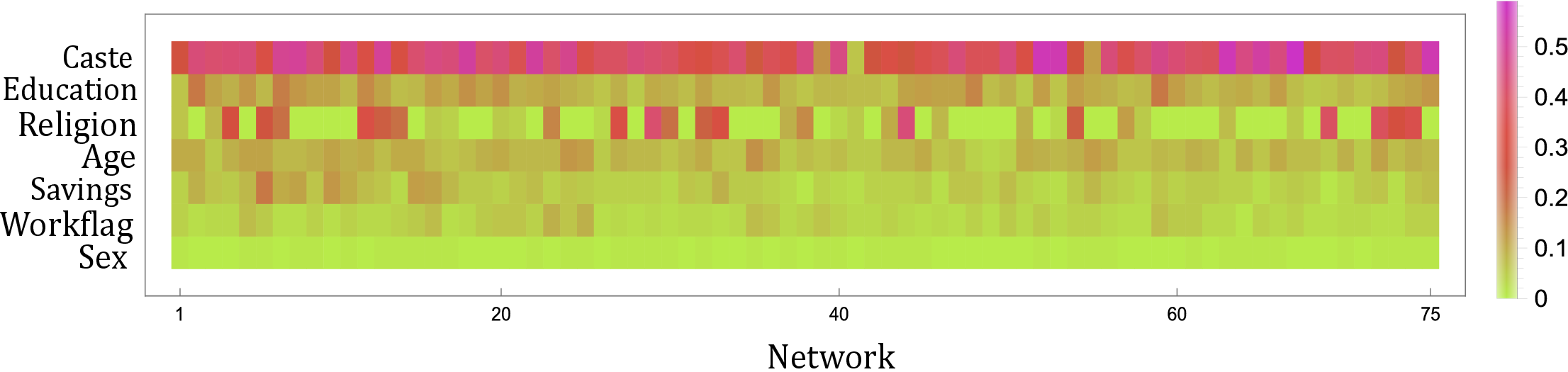}
\caption{Normalized mutual information coefficient between community assignment of nodes and different node covariates for the 75 Karnataka networks. The attributes in the plot, corresponding to different rows, are sorted by the mean normalized mutual information coefficient across the 75 networks.
\label{fig-three}}
\end{figure}

We calculate the normalized mutual information coefficient $\tilde{I}_d$ across all nodes in each village between attribute values and community assignments (Table~\ref{T_two}). We see that caste has a greater $\tilde{I}_d$ (SD) value than the other attributes (caste 0.39 (0.10), sex 0.01 (0.01), age 0.08 (0.02), religion 0.08 (0.12), education 0.10 (0.03), savings 0.05 (0.03), employment 0.04 (0.02)). Considering all networks, The $\tilde{I}_d$ (SD) values for the other attributes are close to 0, showing a weak association with the community assignment (Figure~\ref{fig-three}). This supports the notion that caste is a predictor of network-wide segregation. We also observe that although networks exhibit dyad-level assortativity for the node sex attributes as discussed above, the value of $\tilde{I}_d$ between sex and community assignment is low, suggesting that network communities are more strongly associated with caste than sex. 

\subsection{Assortativity among communities}
To visualize assortativity among communities, we constructed an undirected network of communities. In this community-level network, each node corresponds a community in the individual-level network detected using modularity maximization, and each edge corresponds to the number of ties in the individual-level network that exist between members of the two communities. To simplify visualization, we only included communities that contained at least 5\% of the nodes in the individual-level network in them, and we only included edges between two communities when there existed at least 5\% of the possible connections among the individuals belonging to those communities. The resulting community networks contained on average 77\% of nodes and 26\% of ties present in the underlying individual-level networks.

\begin{figure}[ht]
\centering\includegraphics[width=6in]{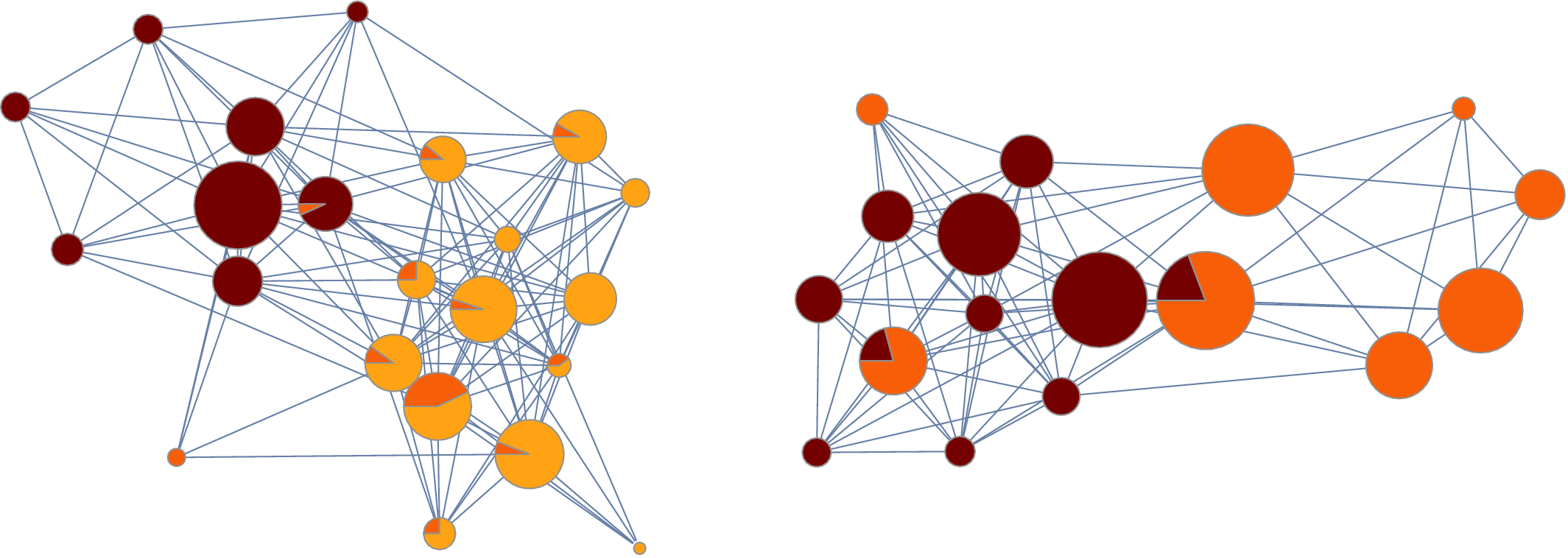}
\caption{Examples of two community-level networks. Each node represents a community detected in the individual-level network using modularity maximization, and each edge represents the existence of one or more ties among individuals in different communities. The community caste distribution is shown as an embedded pie chart within each node (brown: scheduled caste, red: scheduled tribe, orange: OBC, yellow: general).}\label{fig-four}
\end{figure}

We represent the caste distribution for each community as a pie chart embedded in the node (Figure~\ref{fig-four}). Graphically, we observe that communities are mainly composed of people belonging to a single caste. This is consistent with the  $\tilde{I}_d$ values showing that the network community structure can be attributed mainly to the caste of the individuals. In addition, we observe that for some villages, the communities appear to be connected to other communities with similar caste composition (Figures~\ref{fig-four} a,b). 

We adapted the method of modularity maximization ~\cite{Newman2003} to assess if caste-based segregation is also present at the level of communities and not only at the individual level, and we measured caste-based assortativity within and between communities with normalized forms (see Methods). Normalized modularity within communities $Q^*_w$ is positive and higher than 0.3 with a mean (SD) of 0.37 (0.04) in 73 (97.3\%) villages. This result shows that there is strong assortativity based on caste among individuals of the same communities in the majority of villages. On the other hand, modularity between communities $Q^*_b$ is positive for 72 (96.0\%) villages with a mean (SD) of 0.21(0.12). In addition, for 61 (81.3\%) villages $Q^*_b$ is positive and higher than 0.2 showing that there is a tendency of communities to cluster in groups according to their predominant caste. For 11 (14.7\%) of the villages, $Q^*_b$ is close to 0 meaning that there is no segregation between communities driven by caste, and for 3 (4.0\%) of the villages $Q^*_b$ is negative showing a tendency of communities to be related to communities with a different predominant caste. $Q^*_w$ and $Q^*_b$ values for each network are available in the Supplementary Materials.

In resume, we found that, at the dyadic level, same-caste individuals are about 4 times more likely to form a tie with one another than different caste individuals. In addition, we found that there is an association between the networks community partition and the individuals caste. In fact, our results show that same-caste individuals are more likely to form a tie within communities, and that there is a tendency for same-caste communities to connect.

\section{Discussion}
We studied the structure of individual-level and community-level networks in the villages of Karnataka, India. Our main finding is that while every village is relatively densely connected at the individual-level, the networks are segregated at both the individual and community levels. At the dyadic level, we found strong evidence for sex-based assortativity, and in all but one village ties among males occurred more frequently than ties among females, or mixed ties involving males and females. Even as the Indian society is witnessing a shift from being male-dominated ~\cite{Gurvinder2013}, sex-based segregation is still evident in the social networks of the Karnataka villages. This result is consistent with recent findings on emergent inequality in social capital in India where resources and benefits accumulate among different communities and groups of people according to sex and caste.~\cite{Sanyal2015}. Caste continues to play a particular (gender-specific) role in shaping schooling choices of parents increasing the mismatch in education choices and even occupational outcomes between boys and girls in the same caste~\cite{Munshi2006}. From a public health perspective, however, sex-based segregation at the dyadic level does not necessarily exclude one group from the benefits of an intervention targeted at the other. In fact, it has been shown that while men and women report same-sex friendships much more frequently than mixed-sex friendships, mixed-sex ties play an important role in the spread of public health interventions in resource poor settings ~\cite{Hwong2016}. 

We found caste to have a greater effect in segregation than the any other attribute in the study, and segregation by caste occurred in 59 of the 75 villages (78.7\%). This finding supports the notion that caste remains a dominant factor in the discourse on social exclusion in India~\cite{Patil2014}. For all villages, caste was associated with segregation at both the dyadic and community levels. In contrast, sex and other demographics, such as age and employment status, were not associated with network communities. Tie formation and dissolution are often correlated across dyads~\cite{Henry2011}, and here we observed that those behaviors appear to occur at the group level. The effect of segregation depends on the village, and even villages that are geographically nearby can exhibit different levels of caste-based segregation. These differences across villages should be taken into account when planning regional interventions as it has been shown that failure to consider homophily can lead to significant overestimates of the effectiveness of seeding strategies for interventions~\cite{Aral2013}. 
Rural villages need increased attention from public health practitioners given their isolation, vulnerability, low income, and limited access to services. Social markers of inequality are expected to be present virtually everywhere. In India, the caste of a person is an attribute that is both observable and immutable. In general, homophily may be due to multiple mechanisms, social selection and social influence being the two prominent candidate mechanisms. In our study, however, the immutability of caste excludes social influence as a possible mechanism and, furthermore, the observability of caste makes it a plausible target for social selection. These aspects of the caste system make India a compelling country for our study. Even though, the methods presented in this study could contribute to make a similar analysis in contexts where it is not clear which attributes are leading to social segregation. This could contribute to gain awareness of network structure and network effects among policymakers and practitioners enhancing the effectiveness of public health interventions.

This study has some limitations. First, we did not take into account isolated nodes in the analysis, which represented only an average of 2\% of the network nodes. These nodes could be useful for learning about more extreme effects of segregation however we carried out a statistical analysis of the networks structure observing that structural properties, such as the degree distribution and clustering coefficient, were not statistically different when removing the isolated nodes (See Supplementary Materials). Second, attribute data were available for 16983 (25.2\%) nodes and 76440 (26.2\%) of connected node pairs. Because imputation techniques for network data are still in their infancy, we used complete case analysis, i.e., we only included node pairs that had no missing attributes in our models. Consequently, interpretation of model results is subject to this limitation. Finally, the original dataset only included undirected relations and six nodal attributes. It is possible that other attributes that were not measured during the original study, such as political affiliation, are associated with the observed community structure of the networks. Since essentially all network studies, ours certainly included, are observational studies rather than randomized experiments, it is not possible to identify the causes of the observed network structures. More specifically, here it is not possible to identify the causes of the observed network community structure. This is because, using the language of causal inference, all observational studies are subject to unmeasured confounders, where confounding is a bias that arises when the treatment and the outcome share a cause. We therefore stress that the observed community structure is associated with, rather then caused by, certain individual-level attributes.

The statistical model used here is a simple, scalable, and easily interpretable model that belongs to the exponential family. One could alternatively employ different types of models, such as latent space models~\cite{Hoff2002}, stochastic block models ~\cite{Wang1987}, or exponential random graph models (ERGM)~\cite{Robins2007}. We tried fitting a more complicated ERGM to our data, but unfortunately the model failed to converge. Future studies might employ a multiplex network approach using different similarity measures among layers and hence compressing part of the information. This could be useful for detecting differences in the assortativity effects given the different types of relations reported by individuals. Moreover, future studies could benefit from collecting longitudinal data on social networks in the villages and analyze faction and disagreements from a dynamical perspective. This certainly could add more evidence on the causes of segregation among individuals. In fact, new classes of connectivity-informed designs for cluster randomized trials for infectious diseases have been recently proposed, and the designs appear to be able to simultaneously improve public health impact and detect intervention effects \cite{Harling2016}. Adoption of similar designs could improve social and behavioral interventions.
 
\section{Methods}\label{sec_meth}

We used data, now available in the public domain, collected in 75 villages in Karnataka, India, in a study conducted by the Abdul Latif Jameel Poverty Action Lab in 2006 for the Diffusion of Microfinance study published by Banerjee et. al in 2013~\cite{Banerjee2013}. The villages were chosen by Bharatha Swamukti Samsthe (BSS), an organization that operates a conventional group-based microcredit program in India. BSS provided the authors with a list of 75 villages in which they were planning to start operations, and prior to BSS’s entry, these villages had almost no exposure to microfinance institutions, and limited access to any type of formal credit. The data contains information about household-level attributes (e.g., roofing material, type of latrine, quality of access to electric power) and individual-level attributes (e.g., age, sex, religion). The individual-level data was collected in a survey administered only to households that had at least one female aged 18--50 living in the household (about 46\% of the households did). The survey was administered to the head of the household, the spouse of the household head, and to other adult women and their spouses if these women were available for the survey. For non-Hindu households, the survey was administered only if the group represented a minority group in the village, whereas for the Hindu households the survey was randomly administered to 50\% of the households \cite{Banerjee2013}.

We make use of the demographic and social network data collected from surveyed individuals. Demographics included data on sex, age, religion (Hinduism, Islam, Christianity), caste (Scheduled Caste, Scheduled Tribe, Other Backward Class and General), level of education (years), an indicator for whether the individual worked during the week preceding the survey, and an indicator for whether the individual had a bank account. The social network data included the names of people (1) who visit the respondent's home, (2) whose homes the respondent visits, and (3) who are the respondent's kin in the village; it also had the names of people (4) who are relatives with whom the respondent socializes, (5) from whom the respondent receives medical advice, (6) from whom the respondent would borrow money, (7) those to whom the respondent would lend money, (8) those from whom the respondent would borrow material goods (kerosene, rice, etc.), (9) those to whom the respondent would lend material goods, (10) those from whom the respondent gets advice, (11) those to whom the respondent gives advice, and (12) those with whom the respondent goes to pray (at a temple, church, or mosque). Respondents could nominate individuals who did not answer the survey and, consequently, only 16983 (25.2\%) individuals and 76440 (26.2\%) of connected node pairs have demographic information available.

\subsection{Dyad-level Assortativity}
We constructed a simple statistical model to predict the existence of a tie between a pair of nodes based on the sex, age, caste, religion, education, and indicator variables for employment and savings of the individuals. We modeled the binary status of each dyad (tie exists vs. tie does not exist) using logistic regression where similarities of nodal attributes across the dyad were used as predictors. In other words, we considered all node pairs (dyads) in the network, connected or not, and we regressed the binary status of each dyad (0 = not connected, 1 = connected) on the similarity of the corresponding nodal attributes. For continuous attributes, the similarity of two attributes was defined as the difference in their value; for discrete attributes, regardless of the number of categories, the similarity was taken to be 0 if the attributes did not match and 1 if they matched. This model is specified as

\begin{align}\label{eq_one}
\mathbf{P}(Y_{ij}=1|X) = \frac{1}{1+e^{-\beta_0+\sum_{d\in D}\beta_{d}x^{ij}_d}},
\end{align}

\noindent where $Y_{ij}=Y_{ji}$ is 1 if there is a tie between nodes $i$ and $j$, otherwise it is 0. We denote with $D$ the collection of all attributes, and for a given attribute $d$, such as age, $x^{ij}_d$ denote the values of the binary indicator for nodes $i$ and $j$. We use $X$ to denote all model predictors. We estimate this model separately for the LCC of each network, obtaining an estimate and standard error of the regression coefficient $\beta_{d}$ for attribute $d$ in each village. Exponentiated coefficients can be interpreted as odds ratios such that a unit difference in predictor $x_d$ corresponds to a multiplicative change of $e^{\beta_d}$ in the odds.

Since sex-based homophily is common and the logistic model results for dyad-level assortativity by sex do not reflect whether the effect is attributed to male-male or female-female relations, we constructed a mean degree constrained null model for determining whether male-male, female-female and female-male relations were associated with preferential tie formation among individuals. We built the degree constrained null model by keeping the structure of the network fixed and randomly reassigning the sex attribute for each node by a random permutation. We repeated this resampling process 1000 times. Ideally, one would like the null model to preserve correlation between local network properties, like degree and the value of the nodal attribute. The reason this is important is that the number of, say, male-male ties expected under the null model should clearly depend on the number of ties males have as a group. If males have more ties than females, say, an unconstrained permutation of nodal attributes (male, female) will not preserve this observed feature of the data and will lead to a biased comparison between observed and expected tie counts. Here we have chosen to preserve, to a reasonable extent, the mean degree of males and females. We accept any given realization of the null model if the mean degree of males $\langle k^*_{m}\rangle$ and females $\langle k^*_{f}\rangle$ in the permuted networks are within 5\% of their values in the empirical network, denoted by $\langle k^*_{m}\rangle$ and $\langle k^*_{f}\rangle$, respectively. For each such valid simulated network, we count the number of male-male, male-female and female-female ties. Then, we calculate the ratio of observed tie counts to simulated tie counts. After completing 1000 successful simulations, we perform a two-sided test with a Fischer $p$-value for non-symmetrical distributions under the null hypothesis of no sex-based homophily. A $p$-value lower than $0.05$ is interpreted as greater-than-by-chance assortativity.

Demographics are not available for all individuals in the LCCs of the networks, thus we only consider the individuals in the LCC that have data. In the simulated network, males and females may not conserve the structural properties of the original LCC subgraph if we permute the sex attributes among all the LCC nodes (including those with missing attributes). The extent to which this happens can be assessed by comparing the mean degrees of nodes having observed sex attributes with nodes having missing sex attributes. By applying Student's t-test, we observed that structural properties of nodes with missing sex attributes are statistically different from the nodes with observed sex attributes (see Supplementary Materials).

\subsection{Community-level Assortativity}
We assess community-level assortativity by first detecting network communities ~\cite{Kumpula2007,Fortunato2010,Porter2009} and then investigating the extent to which network communities share nodal attribute values. We apply modularity maximization ~\cite{Newman2003,Girvan2002} to detect communities in the LCC of each network using the so-called Louvain heuristic for maximizing modularity~\cite{Blondel2008}. To quantify the association between community membership of nodes ($c$) and nodal attributes ($x_d$), we compute normalized mutual information (NMI)~\cite{Mcdaid2011}

\begin{align}\label{eq_two}
{\tilde{I}}_d= \frac{I(x_d:c)}{\max(H(x_d),H(c))},
\end{align}

\noindent where $I(x_d:c)$ is the mutual information (a non-linear measure of association) between $x_d$ and $c$, $H(x_d)$ is the entropy of $x_d$ and $H(c)$ is the entropy of $c$, where $I(x_d:c)=H(x_d,c)-H(x_d|c)-H(c|x_d)$. $\tilde{I}_d$ is a continuous measure ranging from 0 to 1, and a value of 0 means that the distribution nodal attribute $x_d$ carries no information about the community memberships of nodes, whereas an $\tilde{I}_d$ value of 1 means that node attribute can be mapped to community memberships (Supplementary Materials). Finally, we measure the variability of $\tilde{I}_d$ across the villages in order to see to what extent this association varies across the villages.

\subsubsection{Assortativity within and between Communities}
For a given village, we consider the undirected LCC with $n$ nodes and $m$ edges where each node has been assigned to a single community. The value of modularity $Q$ enables us to quantify assortativity, i.e., to what extent the edges connect nodes of the same type (same attribute value) compared to what we would expect by chance ~\cite{Newman2003}:

\begin{align}\label{eq_three} 
Q= \frac{1}{2m}\sum_{ij}[A_{ij}-\frac{k_ik_j}{2m}] \delta(x_i,x_j).
\end{align}

\noindent Here $A_{ij}$ is the network adjacency matrix, $k_i$ and $k_j$ represent the degrees of nodes $i$ and $j$, $x_i$ and $x_j$ are the attribute values of nodes $i$ and $j$, respectively, and $\delta(x_i,x_j)$ is the Kronecker delta that is equal to 1 if $x_i = x_j$ and 0 otherwise. The modularity $Q$ takes on positive values if there are more edges between nodes sharing an attribute value than what would be expected by chance, and negative values if there are fewer such edges ~\cite{Girvan2002}. Given that modularity maximization assigns every node to a single community, the number of edges $m$ in a given network can be decomposed into two parts, the number of edges between communities $m^b$ and the number of edges within communities $m^w$, yielding $m=m^w+m^b$. Analogous decomposition can be made for node degrees such that the degree $k_i$ of node $i$ is the sum of the number of edges connecting it to nodes of the same community, $k_i^w$, and the number of edges connecting it to nodes in other communities, $k_i^b$, where $k_i = k_i^w + k_i^b$. In order to assess the contribution of edges within communities and edges between communities to modularity, we decompose the modularity measure $Q$ in Eq.~\ref{eq_three} into two measures $Q^w$ and $Q^b$ by including the community assignment of nodes to Eq.~\ref{eq_three}. 

Let $h_i$ represent the community assignment of node $i$ assuming values in the integers $1, \dotsc , H_h$ where $H_h$ is the number of communities detected in the given network. We define the modularity measure $Q^w$ for assessing assortativity based on an attribute of nodes within communities as:

\begin{align}\label{eq_four}
Q^w=\frac{1}{2m^w}\sum_{ij}[A_{ij}-\frac{k^w_ik^w_j}{2m^w}]\delta(x_i,x_j)\delta(h_i,h_j)
\end{align}

\noindent A positive value of $Q^w$ for a given network means that there are more edges between nodes belonging to the same community and having the same attribute value than we expect by chance. 

\noindent Another way to assess assortativity is to investigate if there are more edges between nodes belonging to different communities and having the same attribute value than we expect by chance. This measure would be similar to Eq.~\ref{eq_four} but instead of considering the number of edges $m^w$ that connect nodes within communities and the within-community degree $k_i^w$, we consider the number of edges that connect the nodes across communities $m^b$ and the outside-community degree k$_i^b$ that connect node $i$ to nodes in other communities. The equation for calculating modularity between communities is now given by:

\begin{align}\label{eq_six}
Q^b=\frac{1}{2m^b}\sum_{ij}[A_{ij}-\frac{k^b_ik^b_j}{2m^b}]\delta(x_i,x_j)(1-\delta(h_i,h_j)),
\end{align}

\noindent where the last term of Eq.~\ref{eq_four} is replaced by the 1-complement of the Kronecker delta $\delta(h_i,h_j)$ for only taking into account edges connecting nodes in different communities. A positive value of $Q^b$ for a given network means that there are more edges between nodes belonging to different communities and having the same attribute value than we would expect by chance. 

Across the networks, the maximum value modularity can attain depends on the size of the groups and the degree of the nodes in each network. In order to have comparable results for the different villages, we normalized the within and between community modularities by  dividing the $Q_w$ and $Q_b$ values for a perfectly assortative network~\cite{Newman2010b}.
\begin{align}\label{eq_seven}
Q^*_w=\frac{Q_w}{Q_w^{max}}=\frac{Q_w}{(2m^w-\sum_{ij}(\frac{k^w_ik^w_j}{2m^w})\delta(x_i,x_j)(\delta(h_i,h_j))}
\end{align}
\begin{align}\label{eq_eight}
Q^*_b=\frac{Q_b}{Q_b^{max}}=\frac{Q_b}{(2m^b-\sum_{ij}(\frac{k^b_ik^b_j}{2m^b})\delta(x_i,x_j)(1-\delta(h_i,h_j))}
\end{align}

\section*{Acknowledgments}
We are grateful to R. Zarama, OL. Sarmiento, and the Onnela Lab members for their help at various stages. We acknowledge the Abdul Latif Jameel Poverty Action Lab that has generously placed the data for this research in the public domain.

\bibliography{ReferenceList}
\bibliographystyle{unsrt}

\end{document}